# Hydrogenation of Penta-Graphene Leads to Unexpected Large Improvement in Thermal Conductivity

Xufei Wu,[a] Vikas Varshney,[b,c] Jonghoon Lee,[b,c] Teng Zhang,[a] Jennifer L. Wohlwend,[b,c] Ajit K. Roy,[b] Tengfei Luo [a,d]

a). Aerospace and Mechanical Engineering, University of Notre Dame, Notre Dame, IN 46530

b). Materials and Manufacturing Directorate, Air Force Research Laboratory, Wright-Patterson Air Force Base, OH 45433

c). Universal Technology Corporation, Dayton, OH, 45342

d). Center for Sustainable Energy at Notre Dame, Notre Dame, IN 46530

## Abstract

Penta-graphene (PG) has been identified as a novel 2D material with an intrinsic bandgap, which makes it especially promising for electronics applications. In this work, we use first-principles lattice dynamics and iterative solution of the phonon Boltzmann transport equation (BTE) to determine the thermal conductivity of PG and its more stable derivative – hydrogenated penta-graphene (HPG). As a comparison, we also studied the effect of hydrogenation on graphene thermal conductivity. In contrast to hydrogenation of graphene, which leads to a dramatic decrease in thermal conductivity (from 3590 to 1328 W/mK – a 63% reduction), HPG shows a notable increase in thermal conductivity (615 W/mK), which is 76% higher than that of PG (350 W/mK). The high thermal conductivity of HPG makes it more thermally conductive than most other semi-conducting 2D materials, such as the transition metal chalcogenides. Our detailed analyses show that the primary reason for the counter-intuitive hydrogenation-induced thermal conductivity enhancement is the weaker bond anharmonicity in HPG than PG. This leads to weaker phonon scattering after hydrogenation, despite the increase in the phonon scattering phase space. The high thermal conductivity of HPG may inspire intensive research around HPG and other derivatives of PG as potential materials for future nanoelectronic devices. The fundamental physics understood from this study may open up a new strategy to engineer thermal transport properties of other 2D materials by controlling bond anharmonicity via functionalization.

In the context of nanoelectronics, heat dissipation is considered to be one of the most crucial phenomena. As the electrons flow under applied voltages, current generation leads to dissipative heating, which could be further augmented in 2D material-based nanoelectronics due to the small heat capacity of single layer devices. Such Joule heating problems can eventually affect device performance, reliability, and in turn, shorten their effective lifetime. Hence, it is desired to have highly thermally conductive material components for effective dissipation of the generated heat during device operation.

In the context of materials for nanoelectronics applications, graphene is known to be among the most thermally conductive material in nature with the reported thermal conductivity, $\kappa$, ranging from 1500 to 5000 W/mK.[1-4] Besides its attractive applications,[2,5,6] the unique physics behind the high thermal conductivity is even more fascinating[7-9] – light carbon atoms, strong $sp^2$ bonds and the unusual quadratic out-of-plane phonon modes lead to fast traveling (large group velocity) and hard-to-decay (long lifetime) phonon waves. These not only lead to high thermal conductivity,[1,2,7-9] but research has also shown the divergence in graphene thermal conductivity[7] if there is no boundary limiting how far phonons can travel before they are scattered. Despite its fascinating electrical and thermal properties, graphene is also known to be a semi-metal without a bandgap, which significantly limits its application in transistors.

In graphene, as well as other low dimensional carbon allotropes, such as fullerenes and nanotubes, hexagons (as in honeycomb lattice) serve as the primary building block. Other polygons, such as pentagons, heptagons, *etc.* are considered as topological defects which introduce local bond-frustration in these nanostructures and degrade the electrical,[10] mechanical[11] and thermal[12,13] properties. These defects either occur as single Stone-Wales defect (5, 7 defect) at grain-boundaries or as a pair (5-7-7-5) locally in bulk graphene.

Interestingly, a $C_{20}$ fullerene structure, where pentagons (12 of them) are no longer defects but the fundamental building blocks, was first hypothesized in 2004, and some success has been subsequently attained towards its experimental realization.[14,15] Recently, a new 2D allotrope of carbon based on Cairo-pentagonal tiling pattern, namely, penta-graphene (PG) has been theoretically proposed, which exclusively consists of pentagons in a planar sheet geometry.[16] PG is predicted to be mechanically and dynamically stable up to 1000 K, and more importantly, to have a notable intrinsic band-gap unlike graphene.[16] Possible routes for experimental synthesis of PG were also proposed.[16] Similar to semiconducting transition metal dichalcogenides (TMDs) like $MoS_2$, PG possesses a quasi-direct intrinsic band-gap of ~3.25 eV. In addition, owing to its light mass of carbon atoms along with strong carbon bonds, the thermal transport properties of PG are expected to be high, when compared to TMDs (for $MoS_2$, room temperature thermal conductivity is 18-140 W/mK when a thickness of 6.5 Å is used).[17-23] For example, a recent molecular dynamics simulation, based on an empirical potential, reported the thermal conductivity of PG to be 167 W/mK at room temperature when a thickness of 4.8 Å is used.[24] It is expected that these unique features of PG and its derivatives make them attractive candidates for their incorporation in future nanoelectronics, mechanical, and thermal applications.

The unit cell structure of PG along with its planar geometry is shown in Figure 1. Unlike $sp^2$ hybridized graphene, carbon atoms in PG are of both $sp^2$ (8 atoms/unit cell) and $sp^3$ (4 atoms/unit cell) hybridization.[16] Moreover, in contrast to graphene, the single layer PG consists of 3 atomic planes of carbon atoms (Fig. 1b) – similar to TMDs (such as $MoS_2$) but with different atomic arrangement.[20] In addition, since the surface atoms in PG do not have a symmetrical bonding environment as those in graphene ($sp^2$ carbon atoms are in slightly-non-planar geometry and are thus, under certain bond frustration), they are likely more reactive than those in graphene. In such a case, passivation of $sp^2$ carbon atoms, via hydrogenation for example, is expected to further stabilize the surface layer. For example, hydrogenation of graphene (graphane) is predicted to be more energetically stable than graphene.[25] Hydrogenation is also known to tune other material properties of 2D materials such as electronic and magnetic properties.[26,27] In addition, while graphene is known to be limited in its application in nanoelectronics because of being semi-metal (no bandgap), its hydrogenated form is known to possess a band gap of 3.5 eV.[25]

Although providing a desired band gap, recent calculations on hydrogenated carbon-based materials have shown that the hydrogenation can lead to a significant reduction in thermal conductivity.[8,28] Moreover, if the hydrogenation is only partial, the thermal conductivity reduction can be detrimental due to the hydrogen-induced lattice distortion and thus phonon-defect scattering.[28,29] In the particular case of hydrogenated PG (HPG), as it is more energetically favorable due to the saturation of all non-planar $sp^2$ carbon atoms (hydrogenation in HPG will lead to 18.03 eV lower in energy than PG per primitive cell, see Note 1 in supporting information (SI) for the calculation details), the prediction of thermal conductivity of HPG, as well as its comparison to PG, graphene, and hydrogenated graphene (HG) is important to understand.

In this study, we use first-principles lattice dynamics and iterative solution of the phonon Boltzmann transport equation (BTE) to investigate the thermal conductivity of PG and HPG and put them in perspective with respect to graphene and HG. Interestingly, in contrast to hydrogenation of graphene, which leads to a dramatic decrease in thermal conductivity (from 3590 to 1328 W/mK at a sample length of 62.5 μm), HPG has a thermal conductivity (615 W/mK) 76% higher than that of PG (350 W/mK) at a sample length of 62.5 μm. The high thermal conductivity of HPG makes it more thermally conductive than most other 2D semiconductors, such as the TMDs (for example, $MoS_2$ is reported to have a room temperature thermal conductivity of 18-140 W/mK when a thickness of ~6.5 Å is used.[17-23] ). Our detailed analyses show that the primary reason for the increase in thermal conductivity after hydrogenation is that it makes carbon bonds in HPG less anharmonic than those in PG, despite the increase in phonon scattering channels (*i.e.*, larger phonon scattering phase space). We employ the electron localization function (ELF) as an ideal visualization tool [30,31] for bond characterization and attribute the reason for the relatively better harmonic bond nature in HPG to be complete conversion of C orbitals to *$sp^3$* hybridization upon hydrogenation, leading to pure σ-bonding, which has a more symmetric electron distribution localized between the bonding C atoms. The hydrogenation eliminates the asymmetric distortion of the electron structure related to the distorted π-bond formed by the unhybridized *p* orbitals in the PG bonds.

We use high fidelity first-principles lattice dynamics methodology, which does not require parameterization, to predict the thermal conductivity of graphene, HG, PG and HPG. This method has been repeatedly employed to predict the thermal conductivity of different crystals with great accuracy.[7,8,32-40] To obtain phonon properties, the harmonic force constants are calculated from density functional perturbation theory (DFPT) while the cubic force constants are derived using the finite difference method from a set of force-displacement data obtained from density functional theory (DFT) calculations.[41,42] Phonon group velocity and heat capacity are calculated based on the phonon dispersion relation for each single mode. Employing the cubic force constants, the phonon scattering processes are evaluated by Fermi's Golden rule, and the thermal conductivity is calculated using the iterative solution of the Boltzmann transport equation.[43-45] More detailed description of the calculation is included in the Methods section.

The thermal conductivity of PG and HPG as a function of length is shown in Fig. 2 together with those of graphene and HG. It is worth noting that 2D materials do not have well-defined thicknesses. For graphene, as the interlayer distance of graphite, 3.35 Å,[46] is usually used as the thickness, we choose to use this value for all cases to make a fair comparison among different 2D materials studied here. We have to emphasize the importance of using the same thickness when comparing the thermal conductivity of different 2D materials since all heat has to go through the single layer structures no matter how "thick" or "thin" the structure is. We have further justified this point in detail in Note 2 of SI.

Our calculated thermal conductivity of graphene agrees very well with a recent calculation using the same method as the present work,[47] and it falls in the right range of most experimental results (1500-5500 W/mK at room temperature).[1,48-51] Our values also agree favorably with another first-principles calculation which predicted a thermal conductivity of 3922 W/mK for infinitely long graphene,[8] although our values are smaller than those predicted using the same lattice dynamics method with force constants from Tersoff potential (~3600 W/mK at 10 μm vs. ~2600 W/mK at 10 μm from this work).[7] The differences with the later could be attributed to the empirical nature of the Tersoff potential, which may have a less accurate description of cubic force constants compared to the first-principles DFT calculations.

It can be seen in Fig. 2a (top) that the thermal conductivity of HG is significantly lower than graphene, and the reduction grows to as much as 63% when the sample length reaches 62.5 μm – the largest length studied in this paper. Such a decrease can be easily understood since hydrogenation frustrates the strong carbon bonds in graphene and modifies the planar structure (from atomically smooth to boat or chair confirmation of graphane), leading to reduced phonon group velocities. Surprisingly, the trend is completely opposite for PG after hydrogenation (Fig.

2a (bottom)), which leads to an increase in thermal conductivity by as much as 76%. Again, we emphasize that these thermal conductivity values are calculated using the same thickness of 3.35 Å, which is necessary for a fair comparison between different 2D materials as demonstrated in Note 2 of SI. However, it is worth noting that even if we use different thicknesses as defined in Ref.[24] for the four materials, i.e., the thickness calculated as the summation of the buckling span (the distance between top and bottom atoms) and the van der Waals radii of the outmost atoms, the opposite trends in thermal conductivity change after hydrogenation for graphene vs. PG still exist (see Table S2 in Note 2 of SI), and our analyses and conclusion are still valid.

It is noted that our calculated PG thermal conductivity (350 W/mK) is higher than that from classical MD simulations, which is 240 W/mK corresponding to an adjusted thickness of 3.35 Å used in this study.[24] The difference may be attributed to the inaccuracy of the Tersoff potential especially on the consideration of bond anharmonicity. The thermal conductivity of PG is also larger than 2D TMDs, such as $MoS_2$, which has a room temperature thermal conductivity of 18-140 W/mK when a thickness of ~6.5 Å is used,[17-23] corresponding to 35-270 W/mK when converted with the thickness of 3.35 Å used in the present study.

Besides the total thermal conductivity, we have also calculated the cumulative thermal conductivity as a function of frequency (Fig. 2b), and it is seen that the thermal conductivities of all four materials are dominantly contributed by phonons with frequencies lower than 20 THz. This means that the acoustic phonons dominate the thermal conductivity in all four materials studied here.

To understand the mechanism of the opposite trend of hydrogenation-induced thermal conductivity change for graphene vs. HG and PG vs. HPG, we compare the phonon properties that influence the thermal conductivity. According to the solution to the phonon BTE under the single mode relaxation time approximation (SMRTA), thermal conductivity, $\kappa$, of a 2D material can be expressed as $\kappa = \sum_{p,q} c_{v,pq} v_{pg}^2 \tau_{pg} / 2$,[52] where $c_v$, $v$, and $\tau$ respectively denote the volumetric heat capacity, the phonon group velocity, and the phonon relaxation time, and the subscripts $p$ and $q$ refer to different phonon polarizations and wavevectors. The summation is performed over the whole first Brillouin zone. Strictly speaking, relaxation time is only valid under the SMRTA.[53] Although the thermal conductivity calculation in this work is much more sophisticated, the solution under SMRTA offers a first order approximation for an easy comparison of the relative contributions of different phonon properties towards the thermal conductivity between the studied 2D materials.

Figure 3 shows the phonon dispersion of graphene, HG, PG and HPG. Phonon dispersion curves of both graphene and PG agree well with those from literature (data not shown).[16] For better comparison of phonon frequencies, the maximum frequencies shown are all set to 50 THz whereas the full frequency range dispersion plots are shown in Note 3 of SI. It is obvious that the frequencies of the acoustic modes of HG are much reduced compared to graphene (Fig. 3a and 3b). This leads to a significant reduction in phonon group velocity, which is the slope of the dispersion curves (Fig. 4a (top)). This contributes significantly to the hydrogenation-induced thermal conductivity reduction for HG.

For PG and HPG, the acoustic phonon frequency suppression due to hydrogenation is not obvious (Fig. 3c and 3d). On careful observation, the extracted phonon group velocities are found to be slightly lower for HPG especially at the low frequency range (Fig. 4a (bottom)), while, they are slightly higher than those of PG in 10-20 THz range (circled region in Fig. 4a (bottom)). As the cumulative thermal conductivity plot shows in Fig. 2b (bottom), phonons in the latter frequency range (10-20 THz) have relatively small contributions to the thermal conductivity, it is thus concluded that group velocity cannot be responsible for the increased thermal conductivity of HPG compared to PG.

We further relate the phonon group velocity difference to the atomistic level bonding nature. It is well known that the less "*s*" character a material has, the weaker its bonds are.[54] Thus, we should see that bond strength becomes weaker after hydrogenation, which makes carbon atoms fully $sp^3$ hybridized. As a result, we can roughly sort bond strengths as: graphene (full $sp^2$) > HG (full $sp^3$), and PG (mixture of $sp^2$ and $sp^3$) > HPG (full $sp^3$), and the largest calculated harmonic force constants for each of the four materials shown in Table 1 agree well with this order. So, while the bond strength can be one of the primary reasons of the lower thermal conductivity of HG compared to graphene, it cannot explain the thermal conductivity increase from PG to HPG.

Heat capacity is another factor that influences thermal conductivity. Usually, the specific heat capacity values are not notably different from one material to another. In the context of hydrogenation, it may result in a slightly larger density of materials (since we used the same thickness) and thus higher specific heat capacity. However, this larger specific heat is mainly due to the larger phonon density of states in HG and HPG for high frequency modes. In Fig. 4b, we plot the cumulative specific heat as a function of phonon frequency for all four materials. It is seen that the difference in specific heat is mainly due to modes higher than 20 THz. These modes (> 20 THz) contribute little to the thermal conductivity as shown in Fig. 2b. As a matter of fact, on comparing PG and HPG, we find that the specific heat of HPG is even lower than that of PG up to 20 THz. As a result, the specific heat capacity cannot explain the increased thermal conductivity of HPG compared to PG as well.

If not group velocity and specific heat capacity, differences in phonon scattering processes (such as differences in relaxation times) have to be the governing factor that leads to the large thermal conductivity increase of HPG compared to PG. It is to be noted that the thermal conductivity of HPG under SMRTA is ~ 400 W/mK at 62.5 μm, which is about 1.4 times higher than that of PG (~ 280 W/mK). So, comparing relaxation times shall still offer important insights regarding the mechanism of thermal conductivity difference, despite the fact that the iterative solutions of BTE yield more accurate thermal conductivity values. Figure 4c shows the comparison of relaxation times for each phonon mode as a function of frequency. For graphene vs. HG, it is seen that graphene almost always has larger relaxation times than HG over the whole frequency range, which can also contribute to the larger thermal conductivity of graphene besides the aforementioned group velocity effect.

As the distinction between phonon relaxation times for PG and HPG is not clearly apparent, their comparison is divided into two regions: from 0 to ~4 THz, where the relaxation times of PG are larger than those of HPG, and frequencies larger than 4 THz, where the relaxation times of PG are smaller than those of HPG. To clearly understand how these two regions influence the thermal conductivity difference, we resort back to the cumulative thermal conductivity as a function of frequency as shown in in Fig. 2b (bottom). The figure shows that the dominant contribution to HPG thermal conductivity is from the phonons with frequencies larger than 4 THz, while for PG, phonons with frequencies greater and smaller than 4 THz have similar contributions. As a result, it can be interpreted that the larger relaxation time of phonons with frequencies greater than 4 THz is the primary reason for the higher thermal conductivity of HPG compared to PG.

An interesting finding for HG and HPG is that the relaxation times appear to show a decreasing trend as frequency approaches zero, which is in contrast to the trend of graphene and PG relaxation times. However, on careful observation, the increasing trend is indeed recovered for HG with a dip located around 1-2 THz (Fig. 4c). If observed carefully, a similar dip in graphene relaxation times can also be seen around 2-4 THz. This agrees well with the bump in scattering rate around this frequency range as shown in Ref. [47]. For HPG, it seems that the increasing trend is also emerging but at a much smaller frequency (~ 0.2 THz). Although our calculation could not sample *q* points closer to the gamma point due to limitation on grid size, we believe that at the low frequency limit, the relaxation times of HPG will also increase as seen in HG. However, these extremely low frequency modes are limited by the sample sizes studied in this work and thus should not contribute much to the thermal conductivity. Moreover, thermal conductivity in Fig. 2a (bottom) already shows convergence with respect to length, which further confirm the insignificance of these extremely low frequency modes. This is distinct from graphene, where the very low frequency out-of-plane (ZA) modes can lead to thermal conductivity divergence (Fig. 2a (top)). [7]

We should re-emphasize that the above discussion of relaxation time is based on SMRTA, which includes both Normal and Umklapp processes as resistive scattering mechanisms. However, as momentum is conserved in Normal processes, it should not be treated as resistive scattering mechanism. The iterative solution of BTE takes this factor into account and yields thermal conductivity values 1.5 and 1.3 times the values from SMRTA for HPG and PG, respectively. We further decomposed the phase space into Normal and Umklapp scattering phase spaces (please refer to SI: Note 5), and it was found that the ratio of Normal scattering to Umklapp scattering is similar for these two materials. This is possibly the reason why the ratio of thermal conductivity values from iterative solution and SMRTA are similar for both materials.

The phonon scattering depends on two factors: the likelihood a phonon can be scattered, *i.e.*, how many channels are available for a phonon to get scattered, and the strength of each scattering channel as indicated by the anharmonic force constants. The former factor depends on whether there are three phonon groups that can satisfy the scattering rules (energy and quasi-momentum conservations), and this factor can be quantitatively characterized by the so-called scattering phase space. The latter factor is determined by the anharmonicity of a phonon mode and is usually characterized by the Gruneisen parameter. In order to understand the relative importance of these two factors and to find the root reason of the difference in relaxation times, we compare the three-phonon scattering phase space and the Gruneisen parameter of graphene vs. HG and PG vs. HPG, respectively.

The phase space is calculated as:

$$\frac{2}{3N_q} \sum_{q's',q''s''} \left[ \delta\big(\omega(qs) + \omega(q's') - \omega(q''s'')\big) \delta_{q+q'-q''+G=0} + \frac{1}{2} \delta\big(\omega(qs) - \omega(q's') - \omega(q''s'')\big) \delta_{q-q'-q''+G=0} \right] \quad (1)$$

where $q$, $s$ and $\omega$ refers to the wave vector, branches (*i.e.*, polarization) and angular frequency of phonons. $G$ is the reciprocal lattice parameter, and superscripts refer to different phonons. The delta brackets denote the momentum and energy conservation of three-phonon scattering. $N_q$ refers to the number of sampling $q$-points in the first Brillouin zone. It is to be noted that Eq. (1) is different from the phase space formula defined in Ref. [55], where a normalization factor of $\frac{2}{3N_q^2 N_s^3}$ was used ($N_s$ is the number of phonon branches). We drop the $N_s^3$ factor, which was used for normalization purpose and is not part of the relaxation time formula as shown in Eq. (2), to make sure the comparison is fair for the four 2D materials with different number of phonon branches. We also used $N_q$ instead of $N_q^2$ to make Eq. (1) grid size-independent (please refer to SI: Note 6 for more detailed explanation).

The calculated phase space (Fig. 5a and 5b) shows that hydrogenation leads to increased phase space for both graphene vs. HG and PG vs. HPG. This can be easily understood because adding hydrogen atoms in the primitive cell without changing the symmetry and structure of crystals will increase the amount of phonon branches. These additional phonon branches make satisfying the three-phonon scattering rules easier, and thus phonons can have more channels to get scattered. This can well explain the reason for the shorter relaxation times of HG compared to graphene, since the relaxation time is inversely proportional to the phase space (Eq. (2)).

$$1/\tau_{qs} = 2\pi \sum_{q's',q''s''} \left[ \begin{array}{l} |\tilde{V}_3(qs,q's',-q''s'')|^2 (n_{q's'} - n_{q''s''}) \delta(\omega(qs) + \omega(q's') - \omega(q''s'')) \delta_{q+q'-q''+G=0} \\ + |\tilde{V}_3(-qs,q's',q''s'')|^2 (1 + n_{q's'} + n_{q''s''}) \frac{1}{2} \delta(\omega(qs) - \omega(q's') - \omega(q''s'')) \delta_{q-q'-q''+G=0} \end{array} \right] \quad (2)$$

where $n$ refers to the phonon population, and $\tilde{V}_3$ is the three phonon scattering matrix (please refer to SI: Note 7).

The difference in phase space of PG and HPG, however, cannot explain the larger relaxation times of HPG in the frequency range above 4 THz. Thus, we resort our focus to the anharmonicity (*i.e.*, strength of scattering channels), which is illustrated by the Gruneisen parameter. For graphene and HG (Fig. 5c), phonons below 20 THz have very similar Gruneisen parameters, suggesting that their anharmonicities are similar. On the other hand, HPG shows uniformly smaller Gruneisen parameters than PG (in terms of absolute value) over all frequency range, indicating that HPG is less anharmonic than PG.

The anharmonicity can be indicated by cubic force constants. Table 2 lists the largest cubic force constants of all four materials, which confirms the trend observed in the Gruneisen parameters, *i.e.*, HPG being least anharmonic. As a result, we can conclude that the decrease in bond anharmonicity of HPG due to hydrogenation, which leads to increased phonon scattering strength in the coveted frequency regime (4–20 THz), is the root cause of the increased thermal conductivity of HPG compared to PG.

To further illustrate the bond anharmonicity, we calculated the bond energy profile as a function of bond length. In PG and HPG, carbon bonds can be categorized into two kinds, namely Type I and Type II (see Fig. 6 and Note 8). Type I bonds are the ones between two surface C atoms, and Type II bonds connect the middle layer C atoms with the surface C atoms. We stretch and compress these two types of bonds along the bond direction from their equilibrium bond length and calculate the change in total potential energy (Fig. 6).

It is well-known that the bond anharmonicity is a characterization of the deviation of the bond energy profile from a perfect harmonic (quadratic) profile. Such quadratic fits are presented as dashed lines in Fig 6. In order to quantify the deviation of the actual potential profile from a harmonic one, we calculated the percentile difference between them at +/- 0.1 Å bond length modulation. It is seen that the deviation in HPG bonds are smaller than PG bonds, indicating that HPG bonds are more harmonic. We have also fitted the actual potential profile using a cubic polynomial, where the third order (cubic) constants can also be compared to evaluate relative anharmonicity. It is seen that the PG bonds has larger third order constants than those of HPG bonds (in absolute values), confirming the more harmonic feature of HPG bonds.

To further explore the fundamental relationship between hydrogenation and the observed change in bond anharmonicity, we use first-principles calculations to obtain the ELF, which is defined as:

$$ELF = \frac{1}{1 + \left(D(\mathbf{r})/D_h(\mathbf{r})\right)^2} \tag{3}$$

where $D(\mathbf{r})$ is the curvature of the parallel-spin electron pair density for the actual system at location $\mathbf{r}$, and $D_h(\mathbf{r})$ is the counterpart of a homogeneous electron gas with the same density as that of the actual system at the same location. ELF is a pure ground-state property, which was designed to manifest the bonding features.[30] Furthermore, ELF is dimensionless and allows one to directly compare the bonding between different systems. It has a range from 0 to 1, and ELF=1 means highly localized and bounded electrons, while ELF=0 means lack of electron. The local maxima of these ELF define 'localization attractors', which can be used to illustrate bonding natures.[31]

We have compared the ELF profiles of different types of bonds in PG and HPG (Fig. 7 and Fig. S7 in Note 9 of SI). As seen in Fig. 7, the hydrogenation of the surface C atoms leads to electron localization around the C-H bond (Fig. R3d), while in PG, the electrons around the surface C atoms are more delocalized, seen as the shallower region below the two C atoms (Fig. 7c). It is noted that the surface atoms in PG only bond to 3 atoms and thus have $sp^2$ hybridized orbitals, leaving one $p$ orbital unhybridized. The spatial overlap of the unhybridized $p$ orbitals from the two C atoms form a π-bond, which is indicated by the moon-shaped contour below the two C atoms (Fig. 7c). The π-bond itself does not necessarily lead to larger anharmonicity of PG bonds. However, in PG, the bonds connected to the surface C atoms are not in the same plane, which "squeeze" the $p$ orbital outward – seen as the asymmetric moon-shaped regions above and below the C atoms (Fig. 7c). We can also illustrate this effect by comparing a perfectly planar graphene and its counterpart where two C atoms are displaced out of the plane (see Fig. S8 in Note 10 of SI). It can be seen that the displacement of the two atoms leads to asymmetric bonding environment, which results in the distortion of the $p$ orbital and thus an asymmetric ELF contour for the π-bond. This closely resembles the situation of the Type I bonds in PG (Fig. 7c).

This asymmetric π-bond manifests its influence on bond anharmonicity when the bond is compressed and stretched. When the bond is compressed (Fig. 7e) and stretched (Fig. 7g), the change in electron configuration involved in this π-bond is highly asymmetric, which is evident by comparing Fig. 7e and 7g with 7c. Such asymmetric change in π-bond electron profile should be the root cause of the bond anharmonicity, which is characterized as the deviation of bond energy profile from a perfectly symmetric harmonic well (Fig. 6a).

However, when hydrogenated, the unhybridized $p$ orbital of the C atom hybridizes with the $s$ orbital of the H atoms, making C orbitals $sp^3$ hybridized. The electrons that were previously in the unhybridized $p$ orbitals of the C atoms in PG are now localized at the C-H σ-bond (the big red regions in Fig. 7d) and thus no longer participate in the C-C bonding. As a result, in HPG, the Type I C-C bond is a pure single σ-bond, which is formed by the head-to-head merging of the $sp$ hybridized orbitals from the two C atoms. This bond is featured by the highly localized electron between the two C atoms (Fig. 7d). Compared to its counterpart in PG (Fig. 7c, e, g), the ELF contour of the σ-bond electron in HPG is much more symmetric with respect to the bond axis no matter if the bond is at equilibrium, compressed or stretched (Fig. 7d, f, h). More importantly, compressing and stretching (Fig. 7f and h) of the Type I bond in HPG only change the shape of the electron profile localized at this σ-bond while having no influence on the other orbitals. Such a feature leads to a more elastic C-C bond in HPG as shown in Fig. 6b. This is in contrast to the C-C double bond in PG, where the ELF involved in the π-bond surrounding the C atoms is asymmetrically deformed (Fig. 7e and g).

According to the above discussion, the root reason for the more harmonic bonds in HPG should be that hydrogenation converts the orbitals of the surface C atoms into $sp^3$ hybridization, leading to pure σ-bonding, leading to a more symmetric and localized electron distribution. This eliminates the asymmetric distortion of the ELF related to the distorted π-bond formed by the unhybridized *p* orbitals as seen in the PG bonds. The same observation can be made in the ELF profiles for the Type II bonds, which is shown in Fig. S7 in Note 8 of SI.

It is worth noting that the Type I C-C bond in PG is not a pure π-bond, but a combination of σ-bond and π-bond. The σ-bond is formed by the hybridized $sp^2$ orbitals from the two C atoms, and such a bond is indicated by the more localized electron in-between the two atoms (the red dumbbell shaped region Fig. 7c). This is in contrast to the π-bond formed by the lateral merging of the *p* orbitals (illustrated in Fig. S8 in SI). Although π-bonds are generally weaker than σ-bonds, the superposition of these two bonds leads to a double bond characteristic of the Type I C-C bond in PG, which is stronger than its counterpart in HPG – a pure σ-bond. This is further confirmed by the larger quadratic constant (harmonic force constant) of the fitted cubic polynomials for PG than HPG (Fig. 6).

Finally, we decompose the total thermal conductivity into the contributions from different acoustic modes and optical modes as shown in Fig. S10 in SI. In graphene, the ZA mode (flexural acoustic) dominates thermal conductivity (Fig. S10a). This is consistent with the previous findings from the literature.[7,47] A similar trend was observed for the PG (Fig. S10c). However, on comparing HG to graphene and HPG to PG, we can see that hydrogenation increases the contribution of TA (transverse acoustic) and LA (longitudinal acoustic) modes significantly for both HG and HPG. For HPG, the contribution from the LA mode is almost comparable to that of the ZA mode.

In summary, using first-principles lattice dynamics calculations, we compare the thermal conductivity of graphene, HG, PG and HPG. Hydrogenation, which converts $sp^2$ carbon atoms in graphene and PG into $sp^3$ carbon atoms in HG and HPG, weakens bond strengths and thus decreases phonon group velocities. Although it contributes to the thermal conductivity reduction from graphene to HG, it cannot explain the higher thermal conductivity of HPG compared to PG. It is found that despite the larger scattering phase space, HPG has larger phonon relaxation times compared to PG. The hydrogenation-induced thermal conductivity increase is eventually attributed to the weaker phonon-phonon scattering, which is due to the weaker bond anharmonicity in HPG. The counter-intuitive finding may inspire intensive research around HPG and other derivatives of PG as future nanoelectronics materials and open up a new direction in material engineering to improve thermal conductivity.

## Methods

The potential energy ($V$) and force ($\vec{F}$) of a group of interacting atoms can be expanded using Taylor series expansion with respect to the atomic displacement ($\vec{u}$) when the atoms vibrate around their equilibrium positions (Eqs. (3) and (4), respectively).

$$V = V_0 + \sum_i \Phi_i u_i + \frac{1}{2!}\sum_{i,j} \Psi_{i,j} u_i u_j + \frac{1}{3!}\sum_{i,j,k} \Upsilon_{i,j,k} u_i u_j u_k + ... \quad (3)$$

$$\vec{F}_i = -\frac{\partial V}{\partial \vec{u}_i} = -\Phi_i - \sum_j \Psi_{i,j} \vec{u}_j - \frac{1}{2!}\sum_{j,k} \Upsilon_{i,j,k} \vec{u}_j \vec{u}_k - ... \quad (4)$$

where $\Phi$, $\Psi$, $\Upsilon$ are the first, second (harmonic) and third order (cubic) force constants, respectively, and the subscripts represent indices of atoms. We only considered the force constants up to the third order in our calculations since higher orders anharmonicity have limited effect on phonon scattering unless at very high temperatures.[56] The harmonic force constants are calculated using density functional perturbation theory (DFPT).[57] Non-analytical terms due to the Columbic forces are added to the dynamical matrices[41] with the Born charges and the dielectric constant calculated from DFPT.[57] The cubic force constants are calculated using finite difference according to Eq. (4), where atoms are moved systematically by 0.01 Å from their equilibrium positions and the resultant forces are calculated using density functional theory (DFT) as implemented in Quantum Espresso.[58] The cubic force constants matrices are then constructed from this force-displacement data using the Thirdorder Python tool.[43] The phonon relaxation times and thermal conductivities are calculated using Fermi's Golden rule[59] with the iterative solution of the Boltzmann transport equation using an in-house code.

For all structures studied, ultrasoft pseudopotentials were used with the generalized gradient approximation (GGA)[60,61], parameterized by Perdew-Burke-Ernzerhof,[61] for the exchange-correlation functional. A planewave basis was employed with a cut-off energy of 50 Rydberg and the Monkhorst-Pack[62] scheme was used to generate an 8 x 8 x 1 k-point mesh; both cut-off and mesh size based on the convergence of the lattice energy. Before any force calculation, the atomic structures and the cell sizes are fully relaxed. Harmonic force constants are calculated with a *q*-space grid of 4×4×1, making the effective cutoff of the harmonic force constants larger than 9 Å. The cutoff range of cubic force constants is found to greatly influence thermal conductivity and we used large enough cutoffs to ensure the convergence of the thermal conductivity data (please refer to SI: Note 4). The *q*-space grid of 38×38×1 was used for the Fermi's Golden rule calculation, as this grid size was found to offer well converged thermal conductivity values for all materials studied in this work.

**Acknowledgements**

X.W. thanks the support by NSF (1433490). T.L. thanks the support from the Air Force Summer Faculty Fellowship. The simulations are supported by the Notre Dame Center for Research Computing, and NSF through XSEDE computing resources provided by SDSC Trestles, Comet and TACC Stampede and NCIS Darter under grant number TG-CTS100078.

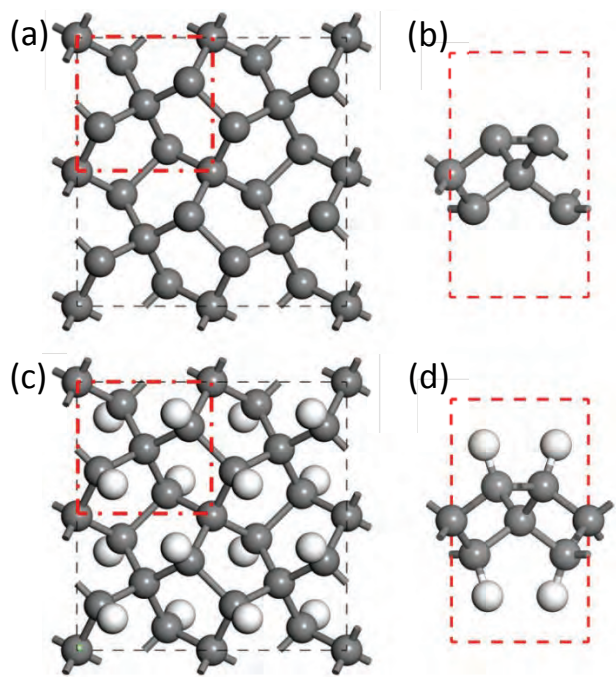

**Figure 1**. Structures of PG and HPG: **(a)** top view and **(b)** side view of PG; **(c)** top view and **(d)** side view of HPG. The red dashed boxes in (a) and (c) indicate the primitive cells.

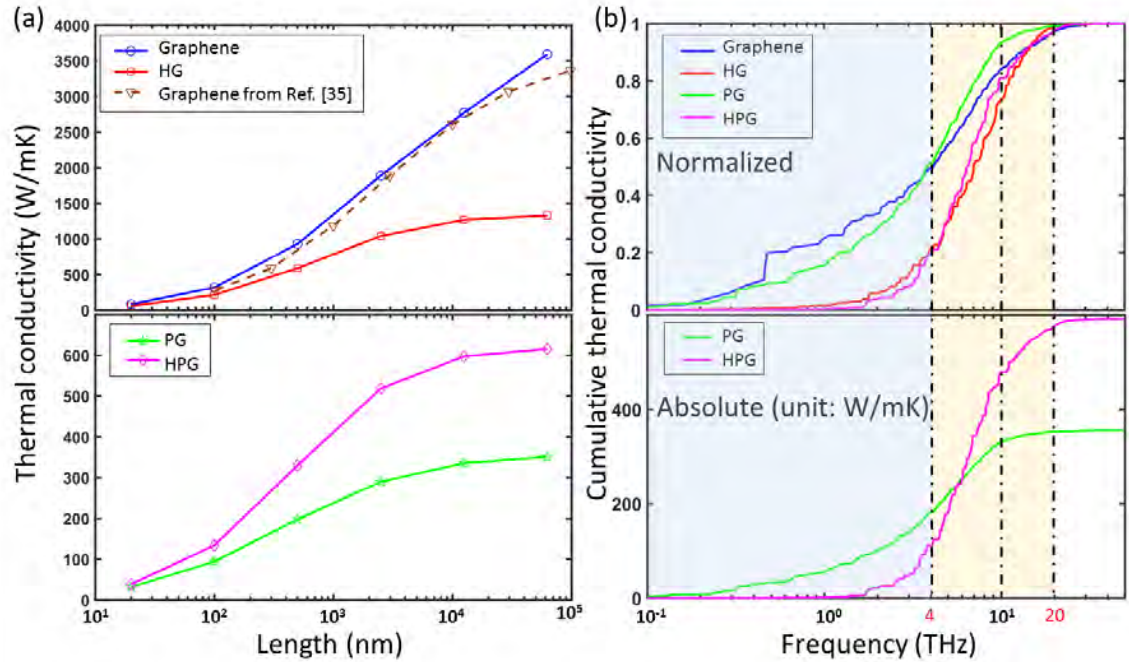

**Figure 2**. **(a)** Thermal conductivity as a function of length: comparison between PG vs. HPG (bottom), and graphene vs. HG (top); **(b)** Normalized cumulative thermal conductivity as a function of frequency for all four materials (top), and absolute cumulative thermal conductivity for PG and HPG (bottom). The frequency is divided in to different ranges as shaded by different colors for better clarity of discussion in appropriate sections of the main text.

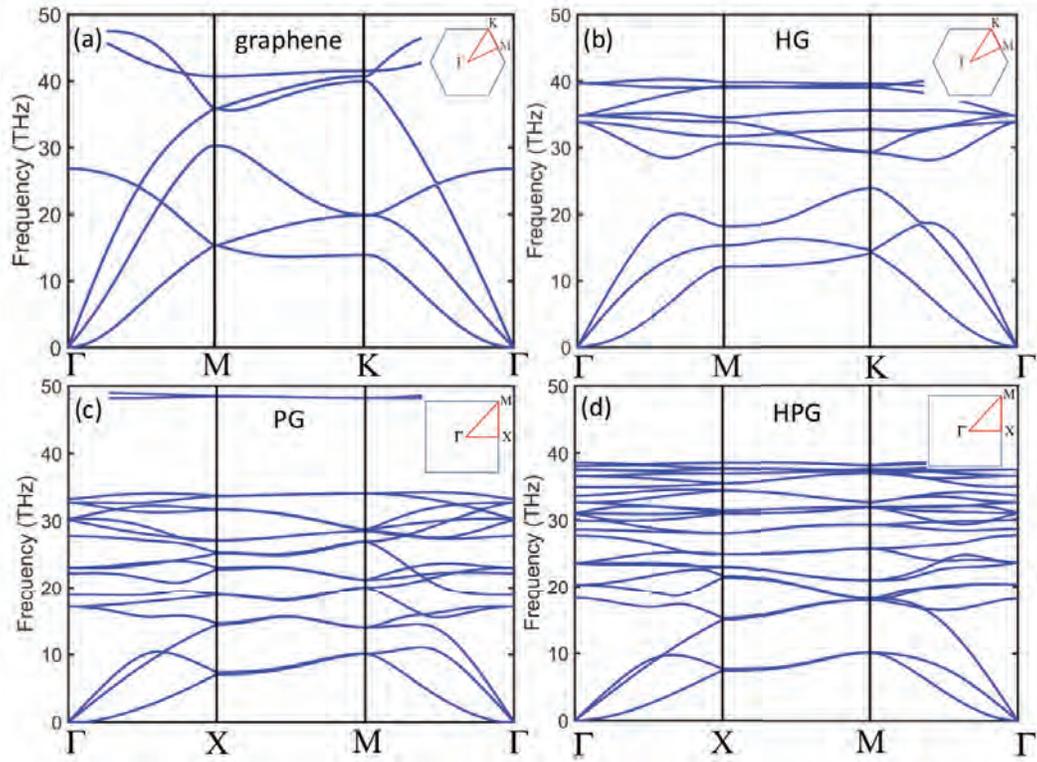

**Figure 3**. Phonon dispersion plots of **(a)** graphene, **(b)** HG, **(c)** PG, and **(d)** HPG. The maximum frequencies for all systems are set to 50 THz for fair comparison of group velocities of different modes. The insets show the first Brillouin Zone and the symmetry lines along which the dispersion curves are plotted.

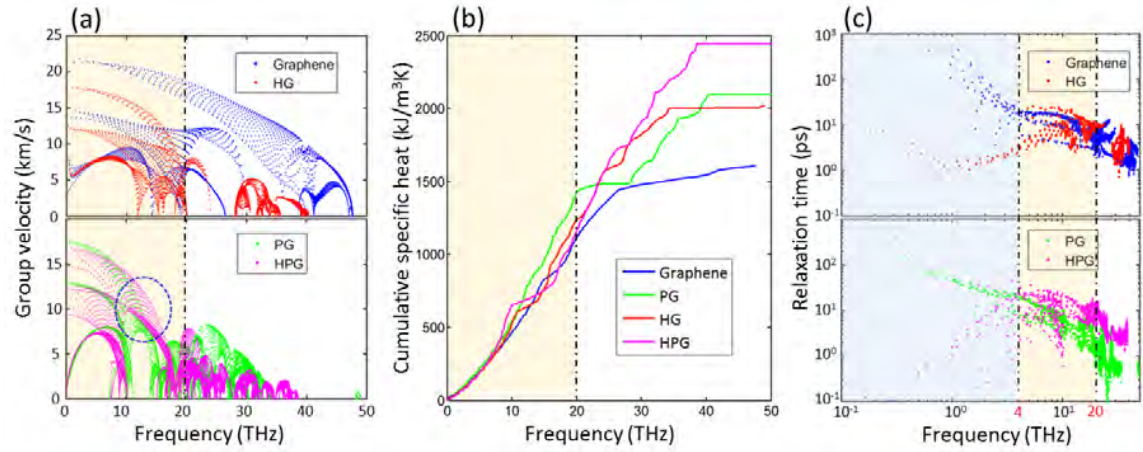

**Figure 4**. Phonon properties: **(a)** comparison of phonon group velocities before and after hydrogenation; **(b)** cumulative specific heat as a function of frequency; **(c)** Phonon relaxation times as calculated via the SMRTA framework.

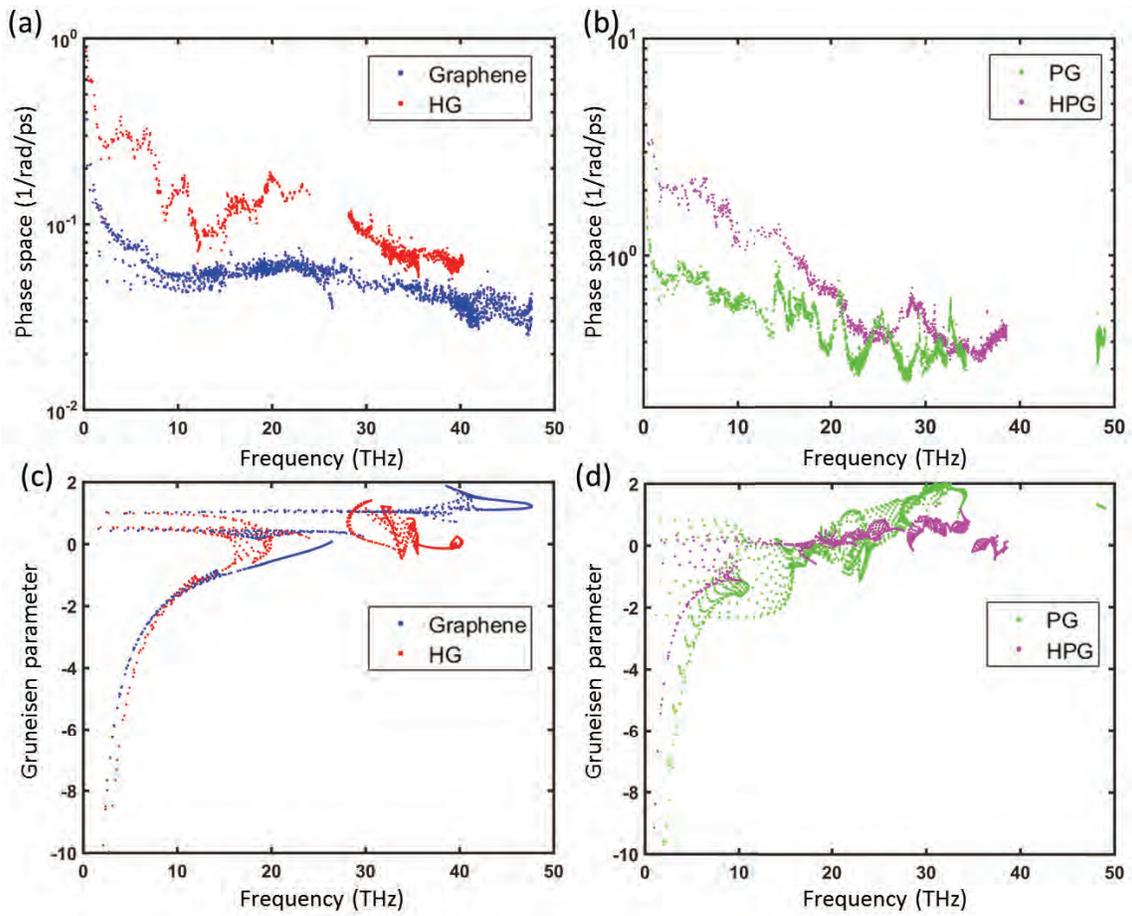

**Figure 5.** Three-phonon scattering phase space of **(a)** graphene and HG; and **(b)** PG and HPG. Gruneisen parameters of **(c)** graphene and HG; and **(d)** PG and HPG.

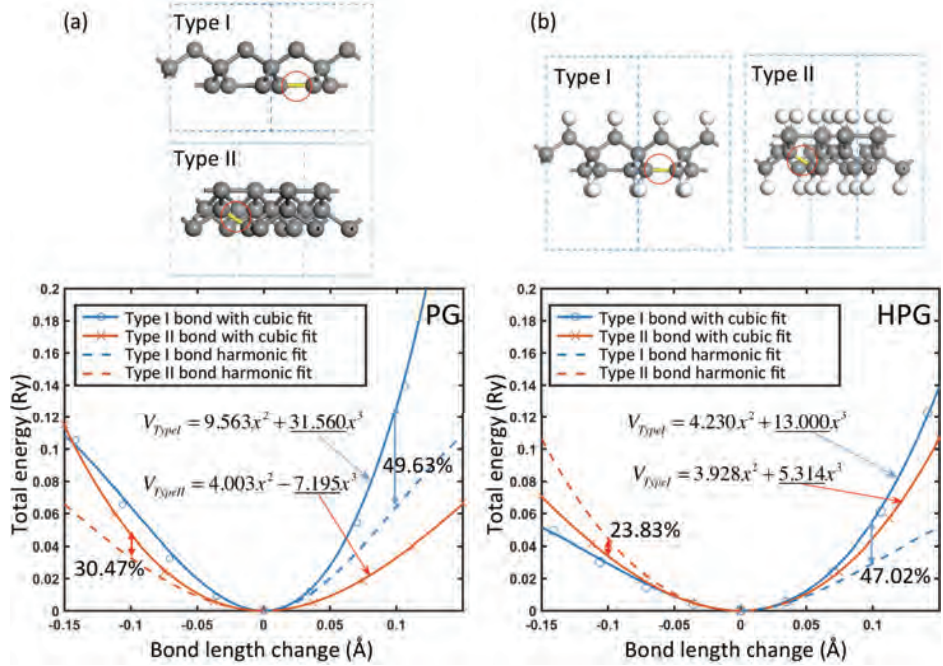

**Figure 6**. Bond energy as a function of bond length modulation for (a) PG and (b) HPG. Both Type I and II bonds are characterized. The solid lines are cubic polynomial fit and the difference in the third order constants (underlined) reflects the relative bond anharmonicity: the larger absolute value of the third order constant indicates larger anharmonicity, i.e., larger deviation from a harmonic profile. The difference between the actual potential profile and an ideal harmonic well (dashed lines) is also characterized by the percentile difference. Bond energy profiles of HPG show smaller deviations from harmonic wells. Note: only half of the harmonic wells (on different sides of minima) are shown for clearer view.

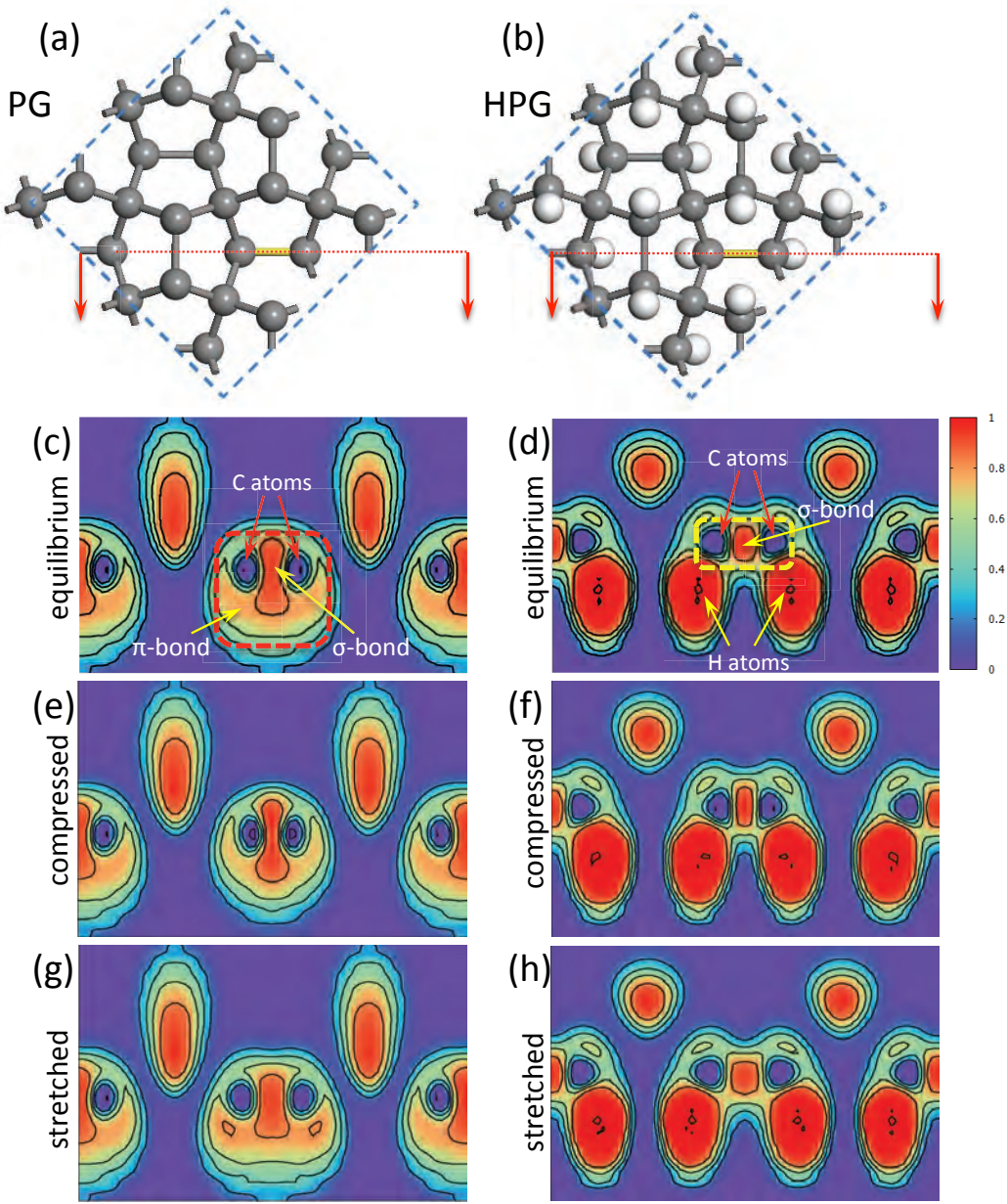

**Figure 7.** ELF of Type I bonds for PG and HPG. (a) and (b) show the atomic structure of PG and HPG and the bonds being studied highlighted in yellow. The red dashed line indicates the location of cross-section, corresponding to which the ELF contours are shown. (c) and (d) are the ELFs for bonds at equilibrium length. (e) and (f) are ELFs for bonds compressed by 0.15 Å, and (g) and (h) are those for bond stretched by 0.15 Å. All ELF contours use the same color scale, which is shown next to panel (d).

**Table 1.** Largest harmonic force constants of the four studied 2D materials

| material | graphene | HG | PG | HPG |
|---|---|---|---|---|
| Largest harmonic force constant (eV/Å$^2$) | 61.049 | 43.239 | 52.549 | 39.753 |

**Table 2.** Largest cubic force constants of the four studied 2D materials

| material | graphene | HG | PG | HPG |
|---|---|---|---|---|
| Largest cubic force constant (eV/Å$^3$) | 184.479 | 202.050 | 137.145 | 122.664 |

Supplementary information for

# Hydrogenation of Penta-Graphene Leads to Unexpected Large Improvement in Thermal Conductivity

Xufei Wu,[a] Vikas Varshney,[b,c] Jonghoon Lee,[b,c] Teng Zhang,[a] Jennifer L. Wohlwend,[b,c] Ajit K. Roy,[b] Tengfei Luo[a,d]

a). Aerospace and Mechanical Engineering, University of Notre Dame, Notre Dame, IN 46530

b). Materials and Manufacturing Directorate, Air Force Research Laboratory, Wright-Patterson Air Force Base, OH 45433

c). Universal Technology Corporation, Dayton, OH, 45342

d). Center for Sustainable Energy at Notre Dame, Notre Dame, IN 46530

## Note 1. Relative Stability of PG and HPG

Standard density functional theory (DFT) calculations, as implemented in Quantum espresso were used to compare the relative stability of PG and HPG, we calculated the hydrogenation energy, which is defined as:

$$E_{hydro} = E_{HPG} - (4 \times E_H + E_{PG}) \quad (S1)$$

where $E_{hydro}$ is the hydrogenation energy, $E_{HPG}$ and $E_{PG}$ are the total energies of the primitive cells of HPG and PG, respectively. $E_H$ is the energy of an isolated hydrogen atom. Ultra-soft pseudopotentials were used with the generalized gradient approximation (GGA), parameterized by Perdew-Burke-Ernzerhof,[1] for the exchange-correlation functional. A planewave basis was employed with a cut-off energy of 50 Rydberg and the Monkhorst-Pack[2] scheme was used to generate an 8 x 8 x 1 k-point mesh; both cut-off energy and mesh size based on the convergence of the lattice energy. $E_{HPG}$ was calculated using the optimized structure and cell size of the HPG primitive cell. $(4 \times E_H + E_{PG})$ is calculated as a whole using the optimized PG primitive cell with 4 hydrogen atoms placed far away from the PG to avoid interaction, and these 4 hydrogen atoms themselves are placed far apart so that they can be treated as isolated atoms. $E_{hydro}$ is calculated to be -18.03 eV per primitive cell.

**Note 2. Justification of using the Same Thickness for 2D Materials in Thermal conductivity Calculations**

In the main text, we have mentioned that for a fair comparison among the different 2D materials studied, we used the same thickness of 3.35 Å for all four materials, which is the interlayer distance of graphite. We argue that employing the same thickness is imperative for comparing the "heat transfer ability" of different 2D materials since all heat has to pass through the single layer materials no matter how "thick" or "thin" they are. We further detail our rationale below:

In 3D materials, the amount of material that participates in heat transfer is proportional to the cross-sectional area (width × height), and this is why when comparing the "heat transfer ability" of different 3D materials, thermal conductivity ($\kappa$) is used, which is defined as:

$$\kappa = -\left(\dot{Q}/(w \times h)\right)/\nabla T \tag{S2}$$

where $\dot{Q}$ is the rate of heat transfer (unit: W), $w$ is the width and $h$ is the height in the cross-sectional plane, and $\nabla T$ is the temperature gradient.

However, in mono-layer 2D materials, no matter how "thick" they are, heat has to pass through the single layer. The amount of material that participates in heat transfer is only proportional to the width (which is clearly defined) of the 2D material, but has nothing to do with the "thickness". As a result, if we were to use a quantity to characterize the "heat transfer ability" of a 2D material, it should ideally be defined as:

$$\kappa' = -\left(\dot{Q}/w\right)/\nabla T \tag{S3}$$

As can be seen, thickness should not be used as a material property when characterizing the heat transfer ability of 2D materials.

We further illustrate such an argument using a thought experiment as shown in Fig. S1. In this figure, there are two 2D materials with different "thicknesses". We assume these two materials have the same width and let them bridge the same heat source and sink. If the rates of heat transfer ($\dot{Q}$) are the same for these two cases, then naturally we say that these two materials have the same "heat transfer ability". We will obtain the same "thermal conductivity" if we use the definition given by Eq. S3, but not if we use Eq. S2 definition.

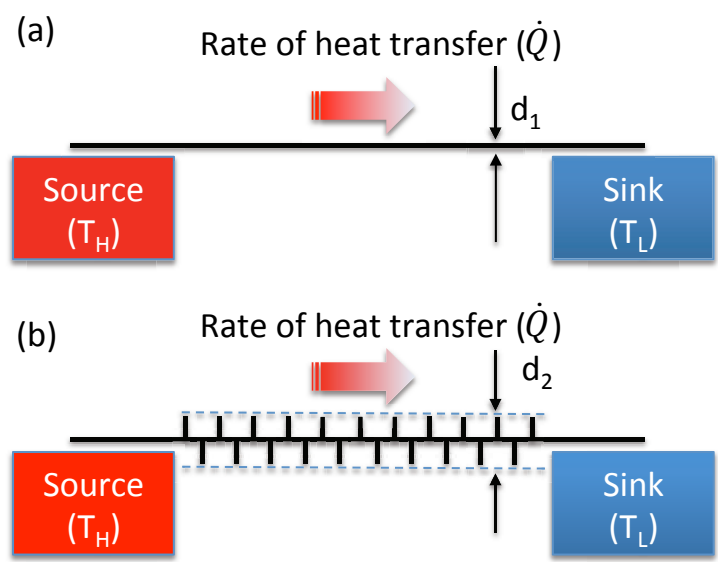

**Figure S1**. Illustration of the heat transfer ability of 2D materials which is independent of thickness.

However, since the "thermal conductivity" defined by Eq. S2 is far more familiar to the scientific community, we still chose to use it in our calculations, but we had to choose a "thickness". According to the above discussion, it is imperative to choose the same thickness when comparing the "heat transfer ability" of different single layer 2D materials to comply with Eq. 2. That is why we chose the interlayer distance of graphite (3.35 Å) as the thickness for all four 2D materials studied.

Researchers have been using the interlayer distance of their 3D counterpart to obtain the nominal thermal conductivity for 2D materials to enable a more sensible comparison between the 2D and the 3D materials. However, when the comparison is made among different 2D materials, the same thickness should be used.

Nevertheless, even if we use different thicknesses for the four 2D materials studied here, the trend does not change, i.e., hydrogenation leads to even larger reduction in thermal conductivity from graphene to HG (by 76%), while the enhancement in thermal conductivity from PG to HPG is still as large as 34%. Such opposite trends are fundamentally interesting and is the focus of this study. We have listed all the thermal conductivity values calculated using the respective thicknesses from buckling distance plus the vdW radius as defined in Ref.[3] in the following table.

**Table S1. Thermal Conductivity Calculated using Different Thicknesses**

| Materials | Thickness (Å) | Thermal conductivity (W/mK) using the same thickness (3.35 Å) | Thermal conductivity (W/mK) using different thicknesses |
|---|---|---|---|
| graphene | 3.35 | 3590 | 3590 |
| HG | 5.08 | 1328 | 876 |
| PG | 4.60 | 350 | 255 |
| HPG | 6.04 | 616 | 342 |

**Note 3. Full Frequency Range Dispersion Relations**

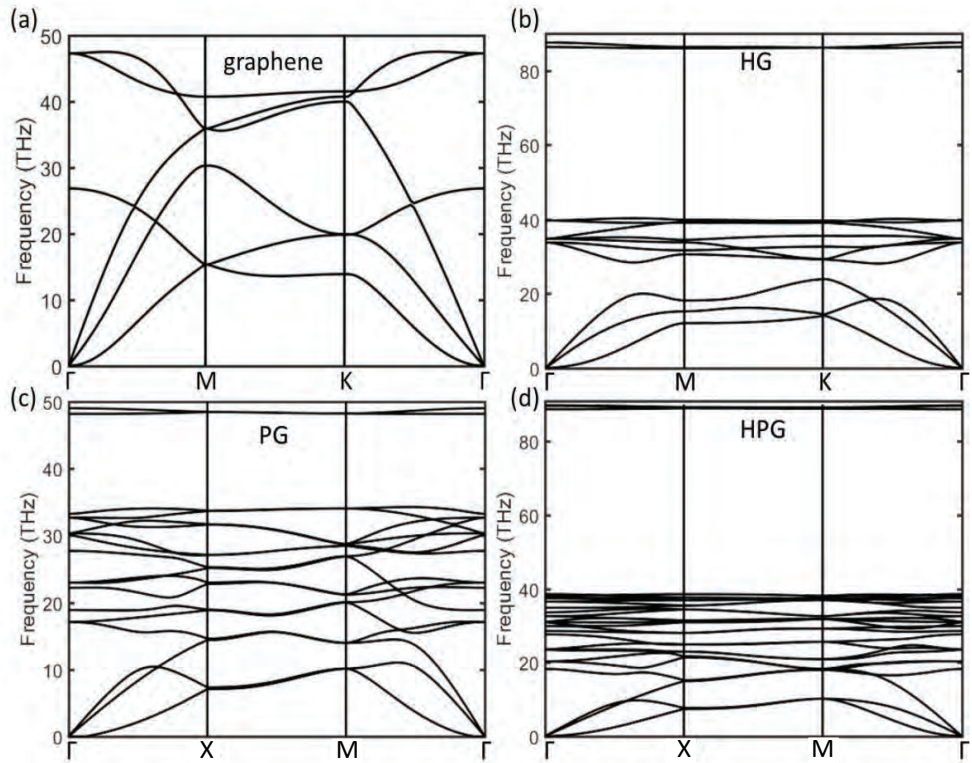

**Figure S2**. Phonon dispersion for (a) graphene, (b) HG, (c) PG, and (d) HPG.

**Note 4. Convergence of Thermal Conductivity**

In order to find out whether we included enough force constants for thermal conductivity calculation, we performed the following test.

For harmonic force constants, the cutoff distance is larger than 9 Å. We tried to include more atoms and compare with the phonon dispersion relation curves. No changes were found on the thermal conductivity when larger cutoff ranges were used.

The calculations for cubic force constants were performed on 4×4×1 supercells. Graphene is well studied and enough literature gives the converged thermal conductivity with a cutoff of third nearest neighbor shell. However, there is no reference for PG, HG and HPG. Here, we plot thermal conductivity as a function of the cutoff distance of the cubic force constants and sample length for PG, HG and HPG.

As can be seen from Fig. S3, for PG, there are two peaks happening at cutoff 4 and 4.5 Å. It is due to the asymmetry of neighbors when these cutoffs are used. We found that the thermal conductivity value drops smoothly again after cutoff 5 Å. For HG and HPG, the convergences are obvious at a cutoff 3.5 Å (Figs. S4 and S5).

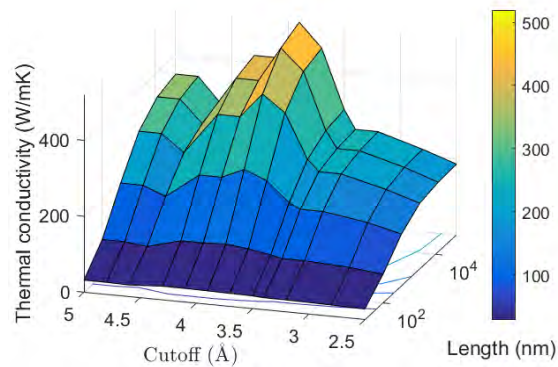

**Figure S3.** Thermal conductivity of PG as a function of material length (nm) and cutoff distance of cubic force constants (Å).

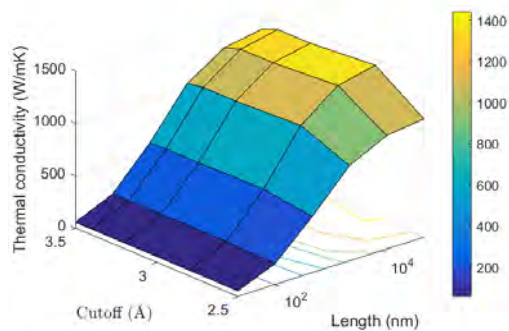

**Figure S4**. Thermal conductivity of HG as a function of material length (nm) and cutoff distance of cubic force constants (Å).

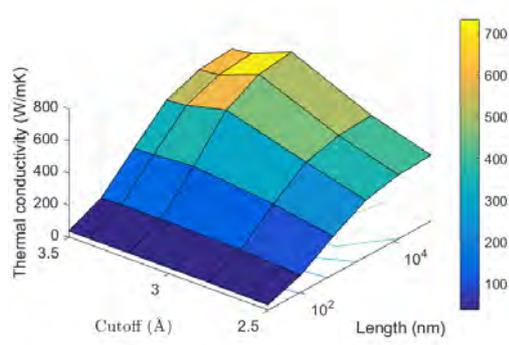

**Figure S5.** Thermal conductivity of HPG as a function of material length (nm) and cutoff distance of cubic force constants (Å).

## Note 5. Decomposition of Total Phase Space for PG and HPG

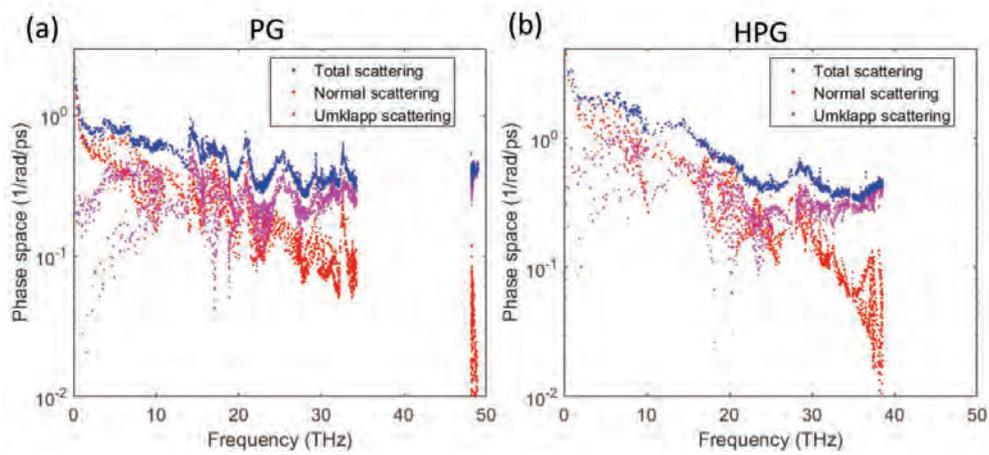

**Figure S6.** Decomposition of three-phonon scattering phase space into Normal and Umklapp scattering for (a) PG and (b) HPG.

## Note 6. Clarification on Phase Space Formula

In Ref [1], the phase space is defined as:

$$\frac{2}{3N_q^2 N_s^3} \sum_{q's',q''s''} \left[ \delta(\omega(qs) + \omega(q's') - \omega(q''s'')) \delta_{q+q'-q''+G=0} + \frac{1}{2} \delta(\omega(qs) - \omega(q's') - \omega(q''s'')) \delta_{q-q'-q''+G=0} \right]$$

(S4)

where $N_q$ is the number of sampling $q$-points in the first Brillouin zone and $N_s$ is the number of phonon branches. $Q$, $s$, and $\omega$ refer to the wave vector, branches and angular frequency of phonons, respectively. The delta brackets denote the momentum and energy conservation of three-phonon scattering. As mentioned in the main text, we have to drop the $N_s^3$ factor, which was used for normalization purpose and is not part of the relaxation time formula (Eq. 2 in the main text).

Furthermore, it turns out that $1/N_q^2$ is not a reasonable pre-factor as it makes the calculated phase space grid size-dependent ($\propto N_q^{-1}$). However, as an intrinsic material property, phonon scattering phase space should not depend on the calculation grid size. If we just consider the plus processes (first part of Eq. S4), a simple proof is:

$$\sum_{q's',q''s''} \delta\big(\omega(qs)+\omega(q's')-\omega(q''s'')\big)\delta_{q+q'-q''+G=0} \propto \sum_{q's',q''s''} \delta_{q+q'-q''+G=0} \propto \sum_{q',q''} \delta_{q+q'-q''+G=0}$$
$$\propto \sum_{q',G} \delta_{q+q'-q''+G=0} \propto \sum_{q'} \propto N_q \qquad (S5)$$

The same goes to the minus processes (second part of Eq. S4). For fair comparison, we modified the phase space equation to

$$\frac{2}{3N_q}\sum_{q's',q''s''}\left[\delta\big(\omega(qs)+\omega(q's')-\omega(q''s'')\big)\delta_{q+q'-q''+G=0}+\frac{1}{2}\delta\big(\omega(qs)-\omega(q's')-\omega(q''s'')\big)\delta_{q-q'-q''+G=0}\right]$$

(S6)

This equation is independent of grid size.

It is worth noting that Eq. (S4) is just another definition and would not cause any problem as the materials have the same number of phonon branches (i.e., same number of atoms in the primitive cell) and the phase spaces are calculated using the same grid size. This was the case in Ref. [4].

**Note 7. Three Phonon Scattering Matrix**

$$\tilde{V}_3(qs,q's',q''s'') = \left(\frac{\hbar}{8N_q\omega(qs)\omega(q's')\omega(q''s'')}\right)^{1/2}\sum_{b0,b'h',b''h''}\sum_{\alpha\beta\gamma}\Phi_{\alpha\beta\gamma}\big(b0,b'h',b''h''\big)$$
$$\times e^{iq'\cdot h'}e^{iq''\cdot h''}\frac{e_\alpha(b|qs)e_\beta(b'|q's')e_\gamma(b''|q''s'')}{\sqrt{m_b m_{b'} m_{b''}}}$$

(S7)

where $\alpha, \beta, \gamma$ represent the *x, y, z* component of Cartesian coordinates, respectively. $\Phi$ is the cubic force constant. *b* indicates the type of atom, and *h* is the translational vector between a specific unit cell and the primitive cell in the supercell.

**Note 8. Two Types of Bonds in PG and HPG**

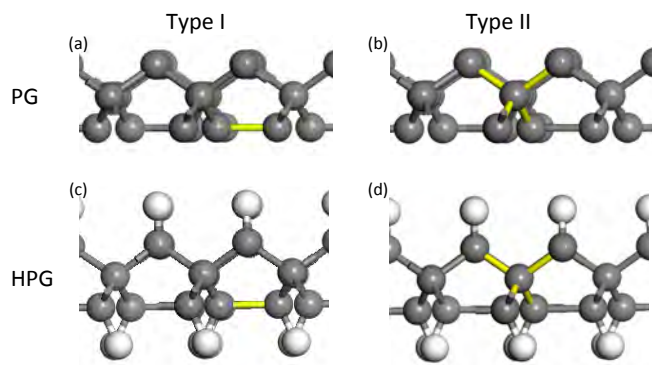

**Figure S7**. Two different types of C-C bonds in PG and HPG highlighted in yellow. (a) and (c) Type I bonds are the ones between two surface C atoms. (b) and (d) Type II bonds connect the middle layer C atoms with the surface C atoms.

**Note 9. ELF Profiles of Type II Bonds in PG and HPG**

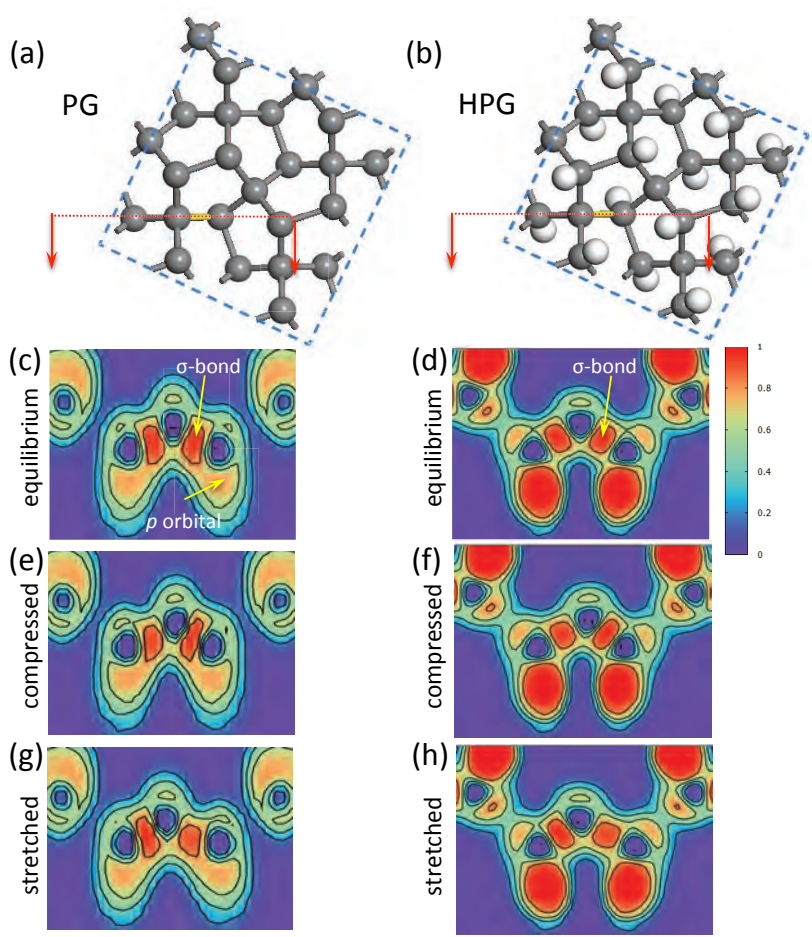

**Figure S8.** ELF of Type II bonds for PG and HPG. (a) and (b) show the atomic structure of PG and HPG and the bonds being studied highlighted in yellow. The red dashed line indicated the location of cross-section, corresponding to which the ELF contours are shown. (c) and (d) are the ELFs for bonds at equilibrium length. (e) and (f) are ELFs for bonds compressed by 0.15 Å, and (g) and (h) are those for bond stretched by 0.15 Å. All ELF contours use the same color scale, which is shown next to panel (d).

**Note 10. Illustration of Distorted *p* Orbitals Leading to Asymmetric π-Bond**

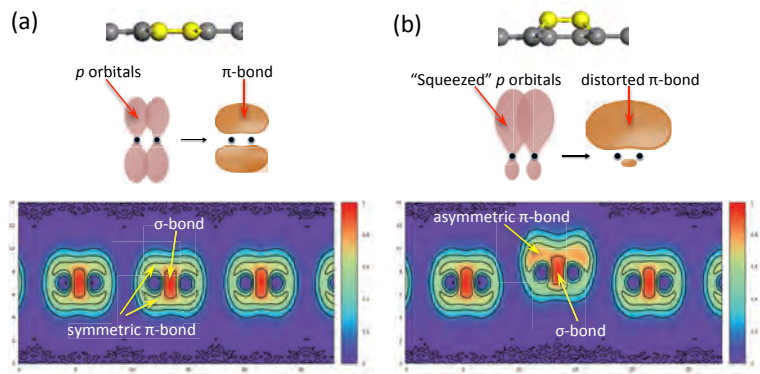

**Figure S9**. Schematics and ELF contours of (a) a symmetric π-bond in planar graphene; and (b) an asymmetric π-bond in distorted graphene. The corresponding ELF contours show a lot similarity with those of the Type I bond in PG.

**Note 11. Decomposition of Thermal Conductivity into Different Phonon Modes**

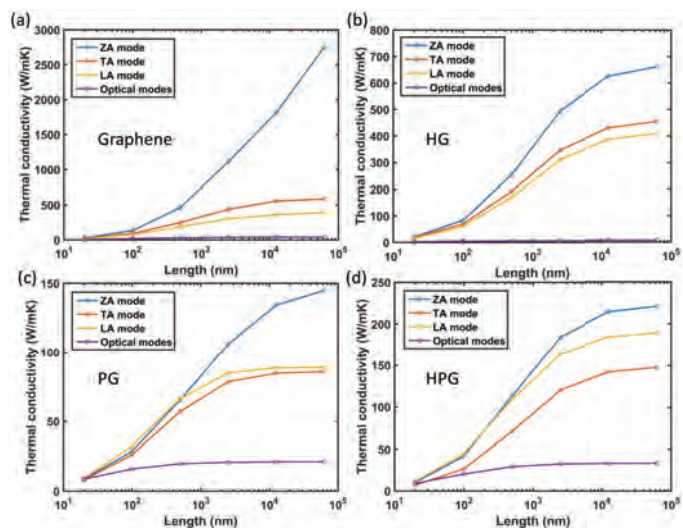

**Figure S10.** Thermal conductivity contribution from different phonon branches for (a) graphene, (b) HG, (c) PG and (d) HPG. ZA – flexural acoustic mode; TA – transverse acoustic mode; LA – longitudinal acoustic mode.